\documentclass[runningheads]{llncs}
\usepackage[T1]{fontenc}
\usepackage{graphicx}

\PassOptionsToPackage{numbers, compress}{natbib}
\usepackage[utf8]{inputenc} % allow utf-8 input
\usepackage{hyperref}       % hyperlinks
\usepackage{url}            % simple URL typesetting
\usepackage{breakurl}
\usepackage{xurl}
\usepackage{multirow}
\usepackage{booktabs}       % professional-quality tables
\usepackage{amsfonts}       % blackboard math symbols
\usepackage{nicefrac}       % compact symbols for 1/2, etc.
\usepackage{microtype}      % microtypography
\usepackage{algorithm}
\usepackage{mathtools}
\usepackage{placeins}
\usepackage{algpseudocode}

\usepackage{bm}

\usepackage{float}
\usepackage{xcolor}

\begin{document}
\title{FinRL-X: An AI-Native Modular Infrastructure for Quantitative Trading}

% \authorrunning{H. Yang et al.}

% \author{Hongyang Yang\thanks{Corresponding author: \email{contact@ai4finance.org}.} \and Boyu Zhang \and Yang She \and Xinyu Liao \and Xiaoli Zhang}
% \institute{
% AI4Finance Foundation\\
% All authors contributed equally to this work.
% }

% \author{Hongyang Yang\thanks{Corresponding author: \email{contact@ai4finance.org}} \and
% Boyu Zhang\and
% Yang She\and
% Xinyu Liao\and
% Xiaoli Zhang}
% \institute{AI4Finance Foundation}

\author{Hongyang Yang\thanks{Corresponding author: \email{contact@ai4finance.org}} \and
Boyu Zhang \and
Yang She \and
Xinyu Liao \and
Xiaoli Zhang}
\institute{AI4Finance Foundation\\
All authors contributed equally.}

\authorrunning{H. Yang et al.}

\maketitle              % typeset the header of the contribution

\vspace{-6mm}
\begin{abstract}

We present FinRL-X, a modular and deployment-consistent trading architecture that unifies data processing, strategy construction, backtesting, and broker execution under a weight-centric interface. While existing open-source platforms are often backtesting- or model-centric, they rarely provide system-level consistency between research evaluation and live deployment. FinRL-X addresses this gap through a composable strategy pipeline that integrates stock selection, portfolio allocation, timing, and portfolio-level risk overlays within a unified protocol. The framework supports both rule-based and AI-driven components, including reinforcement learning allocators and LLM-based sentiment signals, without altering downstream execution semantics. FinRL-X provides an extensible foundation for reproducible, end-to-end quantitative trading research and deployment. The official FinRL-X implementation is available at \href{https://github.com/AI4Finance-Foundation/FinRL-Trading}{https://github.com/AI4Finance-Foundation/FinRL-Trading}.

\end{abstract}

\section{Introduction}

%Quantitative trading research has rapidly progressed in recent years, producing increasingly sophisticated signal models, portfolio construction techniques, and learning-based trading agents \cite{ang2013factor,sahu2023overview,rundo2019machine,zhang2023fingptrag,han2024enhancing}. However, despite methodological advances, many research prototypes remain difficult to reproduce and challenging to deploy. An operational trading system must address a broader set of engineering concerns beyond strategy modeling, including data reliability, interface consistency, execution realism, system robustness, and infrastructure resilience.

Quantitative trading research has rapidly progressed in recent years, producing increasingly sophisticated signal models, portfolio construction techniques, and learning-based trading agents \cite{ang2013factor,sahu2023overview,rundo2019machine,zhang2023fingptrag,han2024enhancing}. However, many research prototypes remain difficult to reproduce and deploy. Practical trading systems must address broader engineering challenges, including data reliability, interface consistency, execution realism, and system robustness.

Existing open-source frameworks typically address isolated stages of the trading pipeline. Recent LLM-based approaches, such as BloombergGPT~\cite{wu2023bloomberggpt}, FinGPT~\cite{yang2023fingpt_open,wang2023fingptbenchmark,liang2024fingpt}, and FinRobot~\cite{yang2024finrobot,zhou2024finrobot}, improve financial text understanding and signal generation, but remain focused on modeling rather than end-to-end system integration. Research-oriented platforms such as FinRL~\cite{liu2020finrl,yang2020deep} and
TensorTrade~\cite{tensortrade} enable rapid experimentation with reinforcement learning agents
but primarily focus on training environments rather than deployment-consistent
architectures. Engineering-oriented libraries including Backtrader~\cite{backtrader}, Zipline~\cite{zipline},
\texttt{bt}~\cite{bt}, vectorbt~\cite{vectorbt}, Qlib~\cite{qlib}, and TradingAgents~\cite{xiao2024tradingagents} provide robust backtesting and evaluation utilities, yet are generally used as standalone components. In
practice, users must still integrate data ingestion, enforce consistent strategy interfaces (e.g., selection–allocation–timing–risk), and implement broker connectivity
and monitoring to obtain a reproducible end-to-end system.

%Existing open-source frameworks typically address only isolated segments of the trading pipeline. Research-oriented platforms such as FinRL~\cite{liu2020finrl,dynamic_datasets,liu2022finrl_meta,liu2021finrl} and TensorTrade~\cite{tensortrade} enable rapid experimentation with reinforcement learning agents but primarily focus on training environments rather than deployment-consistent architectures. Engineering-oriented libraries including Backtrader~\cite{backtrader}, Zipline~\cite{zipline}, \texttt{bt}~\cite{bt}, vectorbt~\cite{vectorbt}, Qlib~\cite{qlib}, and TradingAgents~\cite{xiao2024tradingagents} provide robust backtesting and evaluation utilities, yet are generally used as standalone components. In practice, users must still integrate data ingestion, enforce consistent strategy interfaces (e.g., selection–allocation–timing–risk), and implement broker connectivity and monitoring to obtain a reproducible end-to-end system.

The transition from research backtesting to live deployment introduces system-level distortions that are rarely formalized in academic trading frameworks. We categorize these into two primary deployment gaps.

\textbf{(1) Backtesting-to-paper-trading gap.} 
Offline backtesting environments rely on simplified execution assumptions that diverge from broker-mediated trading environments. Common distortions include oversimplified execution logic (instant fills at bar prices), unrealistic transaction cost modeling, absence of market impact simulation, lack of order book dynamics, survivorship bias, and data feed inconsistencies~\cite{bailey2017probability,algobulls_backtest_vs_live}. These issues create a mismatch between simulated reality and brokered reality, leading to inflated performance metrics and unstable behavior once connected to a trading API. 

\textbf{(2) Paper-trading-to-live-trading gap.} 
Even when strategies pass broker-integrated paper trading, additional execution and operational risks emerge in live markets. These include realistic fill uncertainty (latency, partial fills, slippage), liquidity and queue position effects, API behavior differences, infrastructure fragility (server crashes, disconnections), state recovery failures, real capital constraints (margin rules, settlement timing), and extreme systemic events such as flash crashes or faulty code deployments~\cite{cartea2015algorithmic,alpaca_paper_vs_live}. These factors introduce execution distortion and operational risk that are typically absent in academic simulations.

These two gaps reveal that reproducible modeling alone is insufficient. What is required is a deployment-aware system architecture that preserves interface consistency across research, backtesting, broker simulation, and live execution, while explicitly accounting for execution realism and operational resilience.

%To address these challenges, we introduce \textbf{FinRL-X}, a modular, deployment-oriented trading system built around a unified weight-centric interface. It structures the workflow into four layers: data, strategy (selection–allocation–timing–risk), backtesting, and broker-integrated execution. By preserving consistent weight semantics across layers, FinRL-X reduces discrepancies between offline evaluation and live deployment.

To address these challenges, we introduce \textbf{FinRL-X}, a modular, deployment-oriented trading system built around a unified weight-centric interface. It structures the workflow into four layers: data, strategy, backtesting, and broker-integrated execution, where the strategy layer composes modular decision components. By preserving consistent weight semantics across layers, FinRL-X reduces discrepancies between offline evaluation and live deployment.

%To address these challenges, we introduce \textbf{FinRL-X}, a modular, deployment-oriented quantitative trading system designed around a unified weight-centric interface. As shown in Figure~\ref{overview}, FinRL-X organizes the workflow into four layers: (i) a persistent data layer for reproducible experiments, (ii) a modular strategy layer formalized as selection–allocation–timing–risk transformations, (iii) a standardized backtesting layer built on \texttt{bt}~\cite{bt}, and (iv) a broker-integrated execution layer supporting paper and live trading (e.g., Alpaca~\cite{alpaca_paper_trading}). By preserving consistent weight semantics across layers and incorporating execution monitoring, FinRL-X reduces discrepancies between offline evaluation and deployment behavior.

Our contributions are summarized as follows:
\vspace{-2mm}
\begin{itemize}
\item \textbf{Deployment-aware system architecture.} We formalize and address the backtesting-to-deployment gaps through a layered, weight-centric design that unifies research and execution interfaces.

\item \textbf{Composable trading abstraction.} We structure trading workflows as modular transformations (selection–allocation–timing–risk), enabling seamless integration of rule-based and learning-based strategies without altering downstream components.

\item \textbf{Execution-consistent evaluation.} 
We provide standardized backtesting, broker-integrated execution (e.g., Alpaca~\cite{alpaca_paper_trading}), and monitoring mechanisms to ensure consistency between simulation and deployment.

\item \textbf{Open-source release.} We release FinRL-X as an extensible library with reproducible workflows and runnable examples to facilitate both research and deployment experimentation.
\end{itemize}

\section{Related Work}

Open-source quantitative trading platforms are typically stage-specific: frameworks such as Zipline~\cite{zipline}, Backtrader~\cite{backtrader}, bt~\cite{bt}, and vectorbt~\cite{vectorbt} focus on backtesting, while AI-oriented systems including Qlib~\cite{qlib}, TradingAgents~\cite{xiao2024tradingagents}, and TensorTrade~\cite{tensortrade} emphasize offline ML/RL research. QuantConnect Lean~\cite{quantconnect} offers broker-integrated trading, but is not structured as a modular research-oriented systems architecture. In contrast, FinRL-X adopts a deployment-aware, weight-centric design that unifies data, strategy, evaluation, and execution within a single interface (Table~\ref{tab:platform_comparison}).

\vspace{-5mm}
\begin{table}[htbp]
\centering
\caption{Comparison of FinRL-X with representative open-source quantitative trading platforms.}
\label{tab:platform_comparison}
\resizebox{\linewidth}{!}{
\begin{tabular}{lccccc}
\hline
Feature & FinRL-X & Qlib & TradingAgents & Zipline/Backtrader & QuantConnect Lean \\
\hline
Primary Orientation & End-to-End System & ML Research & Agent-Based Trading & Backtesting & End-to-End Platform \\
Broker Integration  & Yes & No & No & No & Yes \\
Deployment-Consistent Interface & Yes & No & No & No & Partial \\
Reinforcement Learning Support & Yes & Limited & Yes & No & Partial \\
Modular Strategy Pipeline & Yes & No & No & No & Partial \\
Portfolio-Level Risk Overlay & Yes & No & No & No & Partial \\
Open Source License & Apache 2.0 & MIT & Apache 2.0 & Apache 2.0 & Apache 2.0 \\
\hline
\end{tabular}
}
\end{table}
\vspace{-8mm}

\section{Framework}

\begin{figure*}[t]
  \centering
  \makebox[\textwidth][c]{%
    \includegraphics[width=1.05\textwidth]{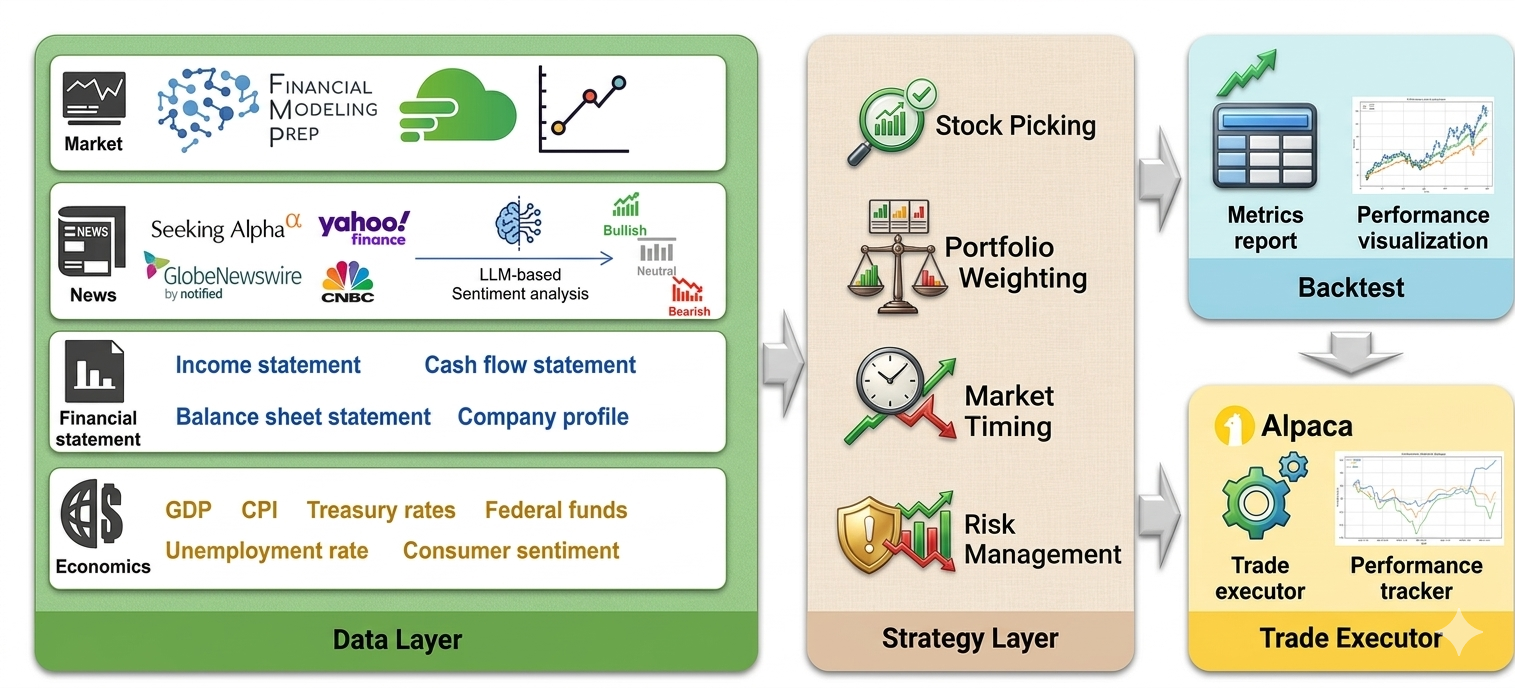}%
  }
  \vspace{-3mm}
   \caption{FinRL-X Framework: A layered, end-to-end trading architecture that unifies data processing, strategy construction, backtesting, and broker-integrated execution within a consistent pipeline, illustrating the workflow from data ingestion to live execution.} 
   \label{overview}
\vspace{-2mm}
  
\end{figure*}

FinRL-X is a modular, deployment-oriented trading platform that structures the quantitative trading workflow into four layers—data, strategy, backtesting, and execution—as shown in Figure~\ref{overview}. Its design goal is to reduce the engineering overhead of building end-to-end systems by enforcing clear module boundaries and stable interfaces, thereby enabling reproducible offline evaluation and seamless transition to paper or live trading.

\subsection{Data Layer}

The data layer provides a unified pipeline for ingesting and normalizing structured (market, fundamental, macro) and unstructured (news) inputs, with primary integration to FMP~\cite{fmp_homepage} and extensible provider support. All sources are aligned to a shared trading calendar to enable consistent rebalancing and evaluation, while news text is transformed into structured sentiment signals via LLM-based preprocessing for integration into the weight-centric strategy pipeline. Reproducibility is ensured through persistent storage of raw snapshots and processed features, reducing discrepancies between offline experiments and deployment.

\subsection{Strategy Layer}

The strategy layer adopts a \emph{weight-centric} architectural principle. 
In FinRL-X, the target portfolio weight vector $w_t \in \mathbb{R}^n$ is treated as the sole interface contract between strategy logic and downstream evaluation or execution modules. 
Rather than emitting trading signals, position deltas, or broker-specific orders, every strategy component produces a target allocation vector that specifies the desired capital fraction assigned to each asset at time $t$.

Formally, let $\mathcal{U}_t$ denote the tradable asset universe at time $t$. 
The strategy layer defines a sequence of contract-preserving transformations that map time-aligned inputs into a feasible portfolio weight vector:
\[
w_t = \mathcal{R}_t\big(\mathcal{T}_t(\mathcal{A}_t(\mathcal{S}_t(\mathcal{X}_{\le t})))\big),
\]
where $\mathcal{S}$ denotes stock selection, $\mathcal{A}$ portfolio allocation, $\mathcal{T}$ timing adjustment, and $\mathcal{R}$ portfolio-level risk overlay. 

This weight-centric abstraction provides three system-level advantages: 
(i) it decouples strategy construction from broker implementation details; 
(ii) it enables composable transformations across heterogeneous rule-based and learning-based modules; and 
(iii) it ensures deployment consistency, as both backtesting and live execution consume the same weight representation.

\begin{algorithm}[htbp]
\caption{Weight-Centric Trading Pipeline}
\label{alg:finrlx_protocol}
\begin{algorithmic}[1]

\Require Data streams $\mathcal{D}, \mathcal{F}, \mathcal{T}, \mathcal{R}$; rebalancing times $\{t_1,\dots,t_n\}$
\State Initialize portfolio value $P_0$

\For{each $t$}
    \State $\mathcal{C}_t \gets \textsc{Select}(\mathcal{F}_{\le t}, \mathcal{U}_t)$
    \State $w_t^{base} \gets \textsc{Allocate}(\mathcal{C}_t)$
    \State $w_t^{timing} \gets \textsc{TimeAdjust}(w_t^{base}, \mathcal{T}_{\le t})$
    \State $w_t \gets \textsc{RiskOverlay}(w_t^{timing}, \mathcal{R}_{\le t})$
    \State Observe realized returns $r_t$
    \State $P_t \gets P_{t-1}(1 + w_t^\top r_t)$
\EndFor

\State \Return $P_n$

\end{algorithmic}
\end{algorithm}
\vspace{-4mm}

\paragraph{Modular Components.}

The pipeline consists of four contract-preserving transformations. 
\textbf{Stock Selection} constructs a candidate set $\mathcal{C}_t \subseteq \mathcal{U}_t$ using fundamentals or learned scoring models under strict no-lookahead semantics. 
\textbf{Portfolio Allocation} maps $\mathcal{C}_t$ to feasible base weights $w_t^{base}$ (e.g., equal-weight, mean--variance, minimum-variance, or DRL-based policies) under consistent normalization and leverage constraints. 
\textbf{Timing Adjustment} transforms $w_t^{base}$ into $w_t^{timing}$ using trend-based or learning-based signals without altering the weight interface. 
\textbf{Risk Overlay} applies volatility-aware exposure scaling (e.g., VIX-based) at the portfolio level, adjusting aggregate exposure while preserving relative allocations to produce final executable weights $w_t$.

\subsection{Backtesting and Execution Layer}
FinRL-X reuses a unified weight interface for both offline backtesting (via \textit{bt}~\cite{bt}) and live broker execution, ensuring consistent portfolio semantics across evaluation and deployment. The executor converts target weights into orders with configurable safeguards and logs realized allocations for post-trade consistency checks.
% \subsection{Deployment-Aware Design}

% Beyond modeling accuracy, quantitative trading systems face systematic distortions when transitioning from research backtesting to live deployment. Let $\mathcal{S}_{research}$, $\mathcal{S}_{paper}$, and $\mathcal{S}_{live}$ denote system behavior under offline simulation, broker-integrated paper trading, and live execution, respectively. In practice,
% \[
% \mathcal{S}_{research} \neq \mathcal{S}_{paper} \neq \mathcal{S}_{live},
% \]
% due to execution simplifications, infrastructure instability, and operational constraints.

% FinRL-X narrows these deployment gaps architecturally. First, it reduces the backtesting-to-paper gap by aligning simulation with broker-style execution semantics, incorporating transaction costs and event-driven order modeling while enforcing consistent data schemas across historical and live ingestion. Second, it mitigates the paper-to-live gap through execution-time safeguards, persistent strategy state for crash recovery, and fault-tolerant broker interaction mechanisms. 

% By preserving a unified weight interface across research and execution layers while engineering for realism and resilience, FinRL-X reduces behavioral divergence between offline evaluation and live deployment.

\subsection{Deployment-Aware Design}

Beyond modeling accuracy, quantitative trading systems face systematic distortions when transitioning from research backtesting to live deployment. Let $\mathcal{S}_{research}$, $\mathcal{S}_{paper}$, and $\mathcal{S}_{live}$ denote system behavior under offline simulation, broker-integrated paper trading, and live execution, respectively. In practice,
\[
\mathcal{S}_{research} \neq \mathcal{S}_{paper} \neq \mathcal{S}_{live},
\]
due to execution simplifications, infrastructure instability, and operational constraints.

FinRL-X narrows these deployment gaps architecturally. It reduces the backtesting-to-paper gap by enforcing consistent execution semantics across environments: strategies output broker-agnostic weight vectors, while simulation incorporates transaction costs, slippage modeling, and event-driven order handling aligned with broker APIs. Data ingestion follows a unified schema to ensure consistency between historical replay and live feeds, minimizing discrepancies caused by data formatting or synchronization differences.

To mitigate the paper-to-live gap, FinRL-X introduces deployment-oriented safeguards at the execution layer. These include state persistence for crash recovery, structured logging for post-trade reconciliation, and fault-tolerant broker interaction mechanisms that handle API interruptions and execution anomalies. Importantly, these mechanisms operate independently of strategy logic, preserving modularity while improving operational resilience.

By maintaining a unified weight interface across research, simulation, and execution layers, and by explicitly engineering for execution realism and robustness, FinRL-X reduces behavioral divergence between offline evaluation and live deployment.

\section{Evaluation}

We evaluate FinRL-X from a system-level perspective, emphasizing reproducibility, modular composability, and deployment consistency in addition to return performance. Experiments compare allocation paradigms, timing mechanisms, and risk overlays under a unified backtesting protocol with standardized metrics.

\subsection{Experimental Setup and Metrics}

Experiments are conducted on liquid U.S. equities and ETFs, with SPY and QQQ as benchmark indices. The historical backtesting horizon spans January 7, 2018 to October 24, 2025 under proportional transaction costs of 10 bps per side. A broker-integrated paper-trading evaluation (e.g., Alpaca~\cite{alpaca_paper_trading}) is conducted from October 26, 2025 to March 12, 2026 to assess deployment behavior. All decisions at time $t$ rely strictly on information available up to $t$, and learning-based models are evaluated using rolling out-of-sample validation.

\textbf{Evaluation metrics.}
Return (cumulative, annualized), risk (volatility, maximum drawdown), risk-adjusted performance (Sharpe, Sortino, Calmar), and deployability (portfolio turnover).

\subsection{Baselines}

We compare FinRL-X with representative baselines from four categories:

\begin{itemize}

\item \textbf{Classical allocation.} 
Equal-weight portfolios serve as a reference baseline. Variance-based portfolio construction methods are also included, namely Mean--Variance optimization~\cite{markowitz1952portfolio} and Minimum-Variance allocation~\cite{clarke2006minimum}.

\item \textbf{Learning-based allocation.} 
Deep reinforcement learning (DRL) allocators generate continuous portfolio weights through sequential decision-making~\cite{mnih2015human,jiang2017deep}. Model selection is performed using rolling out-of-sample validation.

\item \textbf{Timing strategies.} 
Trend-following approaches including Time-Series Momentum (TSMOM)~\cite{moskowitz2012time} and Kaufman Adaptive Moving Average (KAMA)~\cite{kaufman1998trading} provide rule-based market exposure control.

\item \textbf{Risk overlays.} 
A VIX-based volatility scaling mechanism~\cite{whaley2000fear} adjusts portfolio exposure as a modular post-allocation risk management overlay.

\end{itemize}
\vspace{-2mm}

\begin{table}[htbp]
\centering
\caption{Performance and risk metrics across benchmarks}
\label{tab:performance_metrics}
\resizebox{\linewidth}{!}{
\begin{tabular}{lccccccccc}
\hline
Strategy & Cum. Return & Ann. Return & Ann. Vol & Sharpe & Sortino & Calmar & Max DD & DD Duration\\
\hline
SPY                 & 2.63 & 0.14 & 0.17 & 0.84 & 0.76 & 0.60 & -0.23 & 23   \\
QQQ                 & 3.61 & 0.19 & 0.20 & 0.95 & 0.93 & 0.60 & -0.33 & 23 \\
KAMA                 & 2.40 & 0.12 & 0.21 & 0.57 & 0.60 & 0.43 & -0.29 & 23 \\
MeanVar (No Timing) & 2.19 & 0.11 & 0.22 & 0.53 & 0.56 & 0.37 & -0.31 & 41 \\
MeanVar (With Timing)& 2.59 & 0.14 & 0.19 & 0.74 & 0.76 & 0.52 & -0.27 & 38 \\
MinVar (No Timing)  & 2.49 & 0.13 & 0.21 & 0.63 & 0.68 & 0.47 & -0.28 & 25 \\
MinVar (With Timing)& 2.97 & 0.16 & 0.18 & 0.90 & 0.94 & 0.60 & -0.27 & 23 \\
Equal (No Timing)   & 2.11 & 0.11 & 0.20 & 0.54 & 0.53 & 0.40 & -0.27 & 27 \\
Equal (With Timing) & 2.64 & 0.14 & 0.16 & 0.87 & 0.85 & 0.62 & -0.23 & 23 \\
DRL (No Timing)     & 2.33 & 0.12 & 0.22 & 0.55 & 0.54 & 0.40 & -0.31 & 18 \\
DRL (With Timing)   & 3.03 & 0.17 & 0.18 & 0.89 & 0.87 & 0.61 & -0.27 & 18 \\
\hline
\end{tabular}
}
\vspace{-2mm}
\end{table}

\subsection{Portfolio Performance and Ablation Analysis}

FinRL-X is designed to support composable strategy modules under identical execution semantics. We validate this modularity through controlled ablations that isolate timing and overlay effects while keeping the remaining pipeline unchanged.

\begin{figure}[htbp]
    \centering
    \includegraphics[width=1\linewidth]{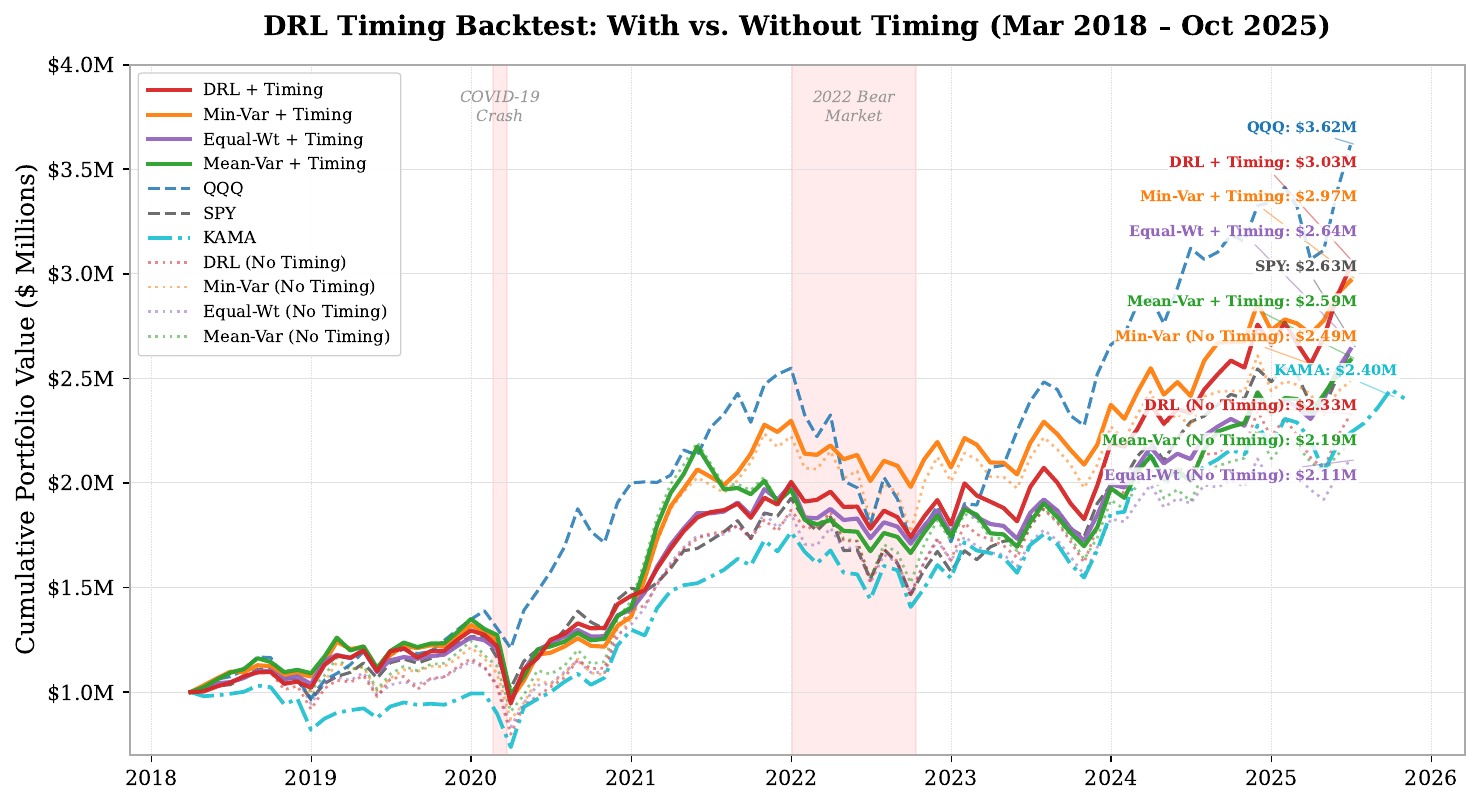}
    \vspace{-4mm}
    \caption{Ablation study of DRL-based allocation with and without timing adjustment. Incorporating the timing module improves cumulative performance and moderates drawdown relative to both the base DRL strategy and the SPY benchmark.}    \label{fig:drl_timing}
    \vspace{-2mm}
\end{figure}

\textbf{Timing ablation (DRL).} Figure~\ref{fig:drl_timing} compares DRL allocation with and without timing against the SPY benchmark. The timing-enhanced variant achieves higher cumulative returns and lower drawdowns, demonstrating that timing can be integrated without modifying backtest or execution interfaces.

\textbf{Cross-strategy ablation.} Table~\ref{tab:performance_metrics} reports standardized return and risk metrics across representative strategies. Across MeanVar, MinVar, Equal, and DRL configurations, timing-enabled variants consistently improve risk-adjusted performance and moderate drawdown relative to their base counterparts.

\label{subsec:ablation}

\subsection{Use Case Demonstrations}
To illustrate end-to-end system flexibility, we present representative use cases that isolate the contribution of individual components while keeping the same workflow (data $\rightarrow$ strategy $\rightarrow$ backtest $\rightarrow$ optional execution) unchanged.

\begin{table}[htbp]
\centering
\caption{Performance comparison of representative use cases and benchmark indices (2018--2025).}
\label{tab:usecase_perf}
\resizebox{\linewidth}{!}{
\begin{tabular}{lcccc}
\hline
Metric 
& Rolling Strategy 
& Adaptive Rotation 
& QQQ 
& SPY \\
\hline
Cumulative Return          & 5.98   & 4.80   & 4.02   & 2.80   \\
Annualized Return (\%)     & 25.85  & 22.32  & 19.56  & 14.14  \\
Annualized Volatility (\%) & 27.85  & 20.30  & 24.20  & 19.61  \\
Sharpe Ratio               & 0.93   & 1.10   & 0.81   & 0.72   \\
Maximum Drawdown (\%)      & -38.95 & -21.46 & -35.12 & -33.72 \\
Calmar Ratio               & 0.66   & 1.04   & 0.56   & 0.42   \\
Win Rate (\%)              & 54.36  & 54.77  & 56.25  & 55.28  \\
\hline
\end{tabular}
}
\end{table}

\paragraph{Use Case 1: Portfolio Allocation Paradigms}

We evaluate heterogeneous portfolio allocation mechanisms under a unified weight-centric interface, including learning-based DRL allocation, classical optimization-based methods (Mean--Variance, Minimum-Variance), equal-weight baselines, and signal-driven timing strategies such as KAMA operating as standalone weighting pathways.
By enforcing identical data and execution semantics, FinRL-X enables fair comparison across fundamentally different allocation paradigms without architectural modification.
Backtesting results for this use case are consolidated in Figure~\ref{fig:drl_timing} and Table~\ref{tab:performance_metrics}.

\vspace{-6mm}
\begin{figure}[htbp]
    \centering
    \includegraphics[width=1\linewidth]{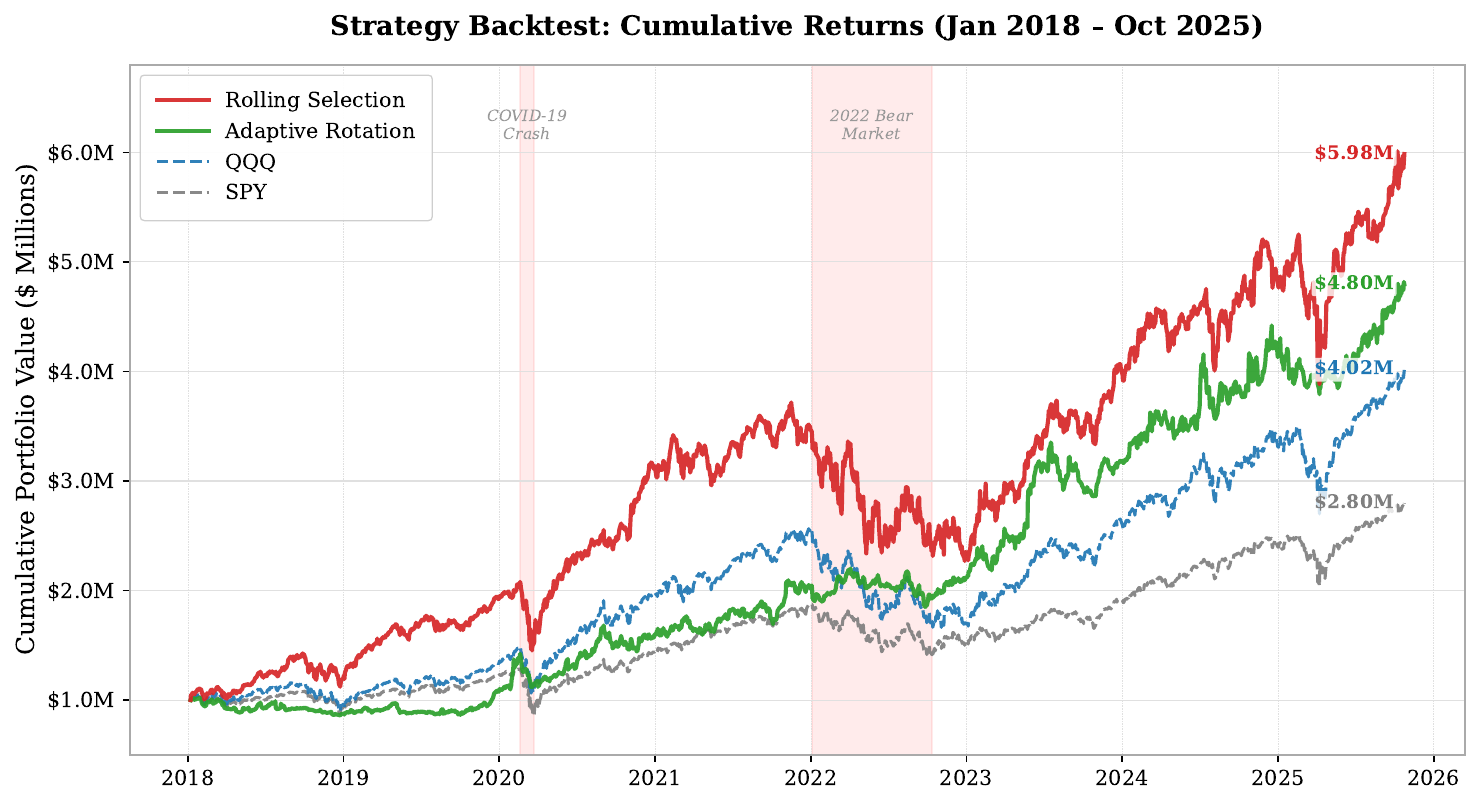}
    \vspace{-4mm}
    \caption{Backtest performance comparison across representative strategy configurations under the unified weight-centric protocol (January 7, 2018 – October 24, 2025). Results illustrate cumulative portfolio trajectories relative to benchmark references.}
    \label{fig:All_Backtests}
\end{figure}
\vspace{-6mm}

\paragraph{Use Case 2: Rolling Stock Selection}
This use case tests the rolling stock selection module, where the universe is updated upon new quarterly financial reports. We use all component stocks of the NASDAQ 100 index as candidates and select the top 25\% to construct a portfolio with DRL-based allocation. Figure~\ref{fig:All_Backtests} (line \textbf{Rolling Selection}) shows cumulative returns. Table~\ref{tab:usecase_perf} (Rolling Strategy) reports performance relative to SPY and QQQ.

\paragraph{Use Case 3: Adaptive Multi-Asset Rotation}
This use case presents an adaptive multi-asset rotation strategy designed to achieve stable excess returns relative to QQQ across regimes. Assets are grouped into Growth, Real Assets, and Defensive buckets, with at most two active groups selected per weekly rebalance. Group selection is driven by Information Ratio relative to QQQ, while intra-group allocation uses residual momentum with robust exception handling. Regime indicators are used for risk gating rather than alpha generation.

Figure~\ref{fig:All_Backtests} (line \textbf{Adaptive Rotation}) shows sustained outperformance with improved drawdown control across cycles. Table~\ref{tab:usecase_perf} (Adaptive Rotation) reports risk-adjusted metrics and drawdown improvements relative to SPY and QQQ.

\subsection{Paper Trading and Deployment Validation}

\vspace{-4mm}
\begin{figure}[htbp]
    \centering
    \includegraphics[width=1\linewidth]{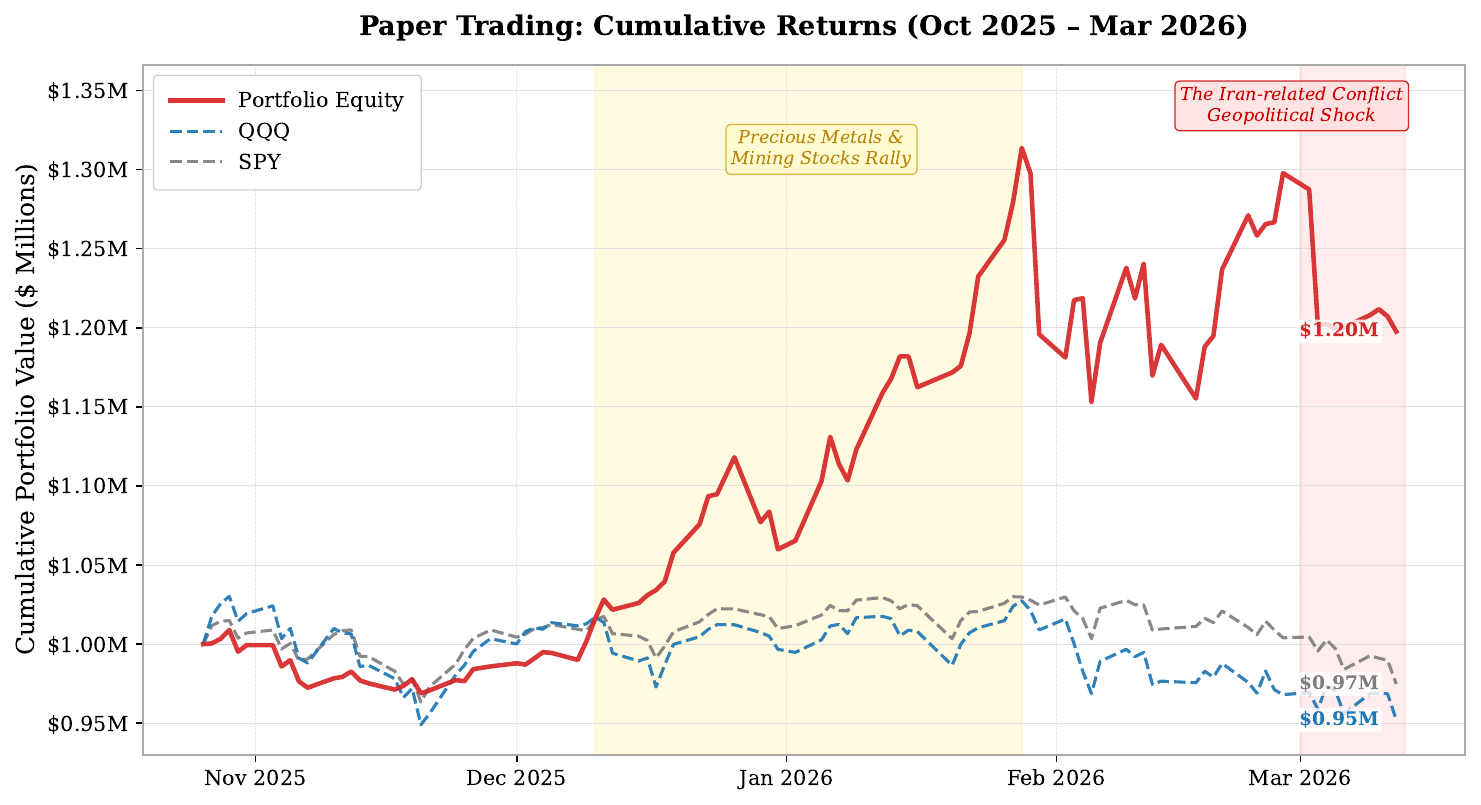}
    \vspace{-4mm}
    \caption{Paper trading performance relative to benchmark indices (October 26, 2025 -- March 12, 2026), demonstrating deployment-consistent execution under daily rebalancing.}    
    \label{fig:Paper_Trading}
    \vspace{-4mm}
\end{figure}

\textbf{Paper trading as deployment-consistency validation.}
To bridge offline evaluation and live deployment, we execute an ensemble strategy combining Rolling Selection and Adaptive Rotation in an Alpaca paper trading environment from October 2025 to March 2026 under daily rebalancing. While the evaluation horizon is limited, the results demonstrate stable deployment behavior and consistent execution under real broker conditions. Specifically, the experiment serves to validate operational robustness and consistency between offline portfolio targets and broker-level execution. Figure~\ref{fig:Paper_Trading} and Table~\ref{tab:paper_trading} present the resulting equity curve and summary performance statistics.

\begin{table}[htbp]
\centering
\caption{Performance comparison between paper trading and benchmark indices (Oct 26, 2025--Mar 12, 2026, Daily Turnover).}
\label{tab:paper_trading}
\begin{tabular}{lccc}
\hline
\textbf{Metric} & \textbf{Strategy} & \textbf{SPY} & \textbf{QQQ} \\
\hline
Cumulative Return        & 1.20 & 0.97 & 0.95 \\
Total Return (\%)        & 19.76 & -2.51 & -4.79 \\
Annualized Return (\%)  & 62.16  & -6.60  & -12.32  \\
Annualized Volatility (\%) & 31.75  & 11.96 & 16.79 \\
Sharpe Ratio             & 1.96   & -0.55   & -0.73   \\
Maximum Drawdown (\%)    & -12.22 & -5.35 & -7.88 \\
Calmar Ratio             & 5.09   & -1.23   & -1.56   \\
Win Rate (\%)            & 64.89  & 52.13  & 54.02 \\
\hline
\end{tabular}
\end{table}
\vspace{-4mm}

In addition to return metrics, we track deployment-oriented indicators such as 
order rejection rate, execution guardrail triggers, and portfolio weight tracking 
error between target and realized allocations. These indicators remain consistently 
low throughout the paper trading period, suggesting stable execution behavior and 
high fidelity between target and realized portfolios.

\vspace{-4mm}
\begin{figure}[htbp]
    \centering
    \includegraphics[width=0.8\linewidth]{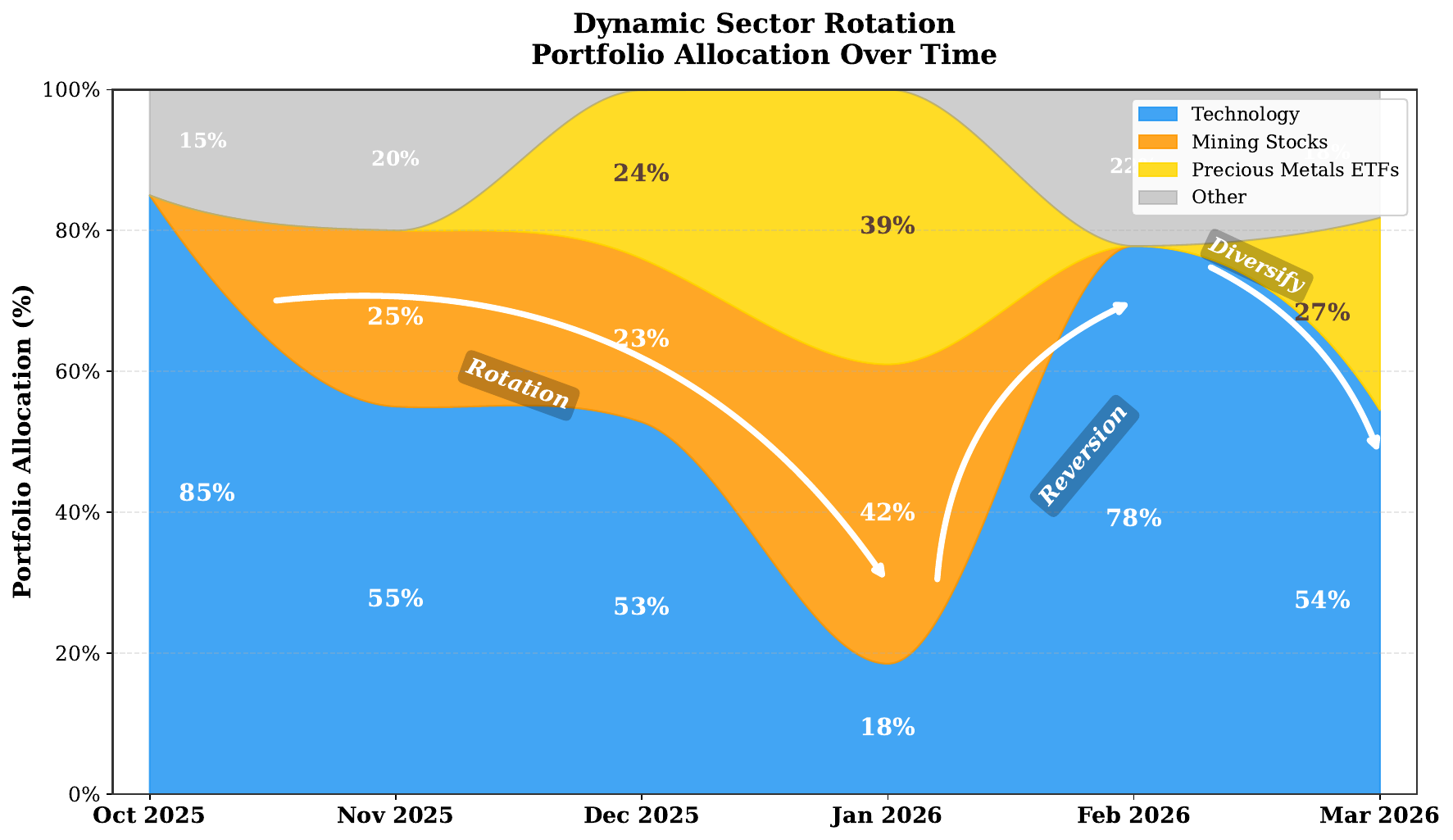}
    \vspace{-2mm}
    \caption{Portfolio allocation trajectory under the unified weight-based execution framework during paper trading. The figure illustrates time-varying exposure adjustments across asset groups, demonstrating modular allocation outputs that are directly executable without architectural changes.}
    \label{fig:Allocation_Trajectory}
    \vspace{-5mm}
\end{figure}
\vspace{-6mm}

\subsubsection{Paper Trading Analysis}

To evaluate deployment consistency beyond offline backtesting, we conducted a six-month paper trading session from October 26, 2025 to March 12, 2026 using the ensemble configuration under daily rebalancing.

As shown in Figure~\ref{fig:Paper_Trading}, the strategy achieved a total return of \textbf{+19.76\%}, outperforming both SPY and QQQ over the same period. Given the limited horizon, these results are not intended to establish statistically significant alpha. Rather, the experiment validates the end-to-end execution pipeline, including portfolio generation, broker connectivity, order routing, execution monitoring, and post-trade reconciliation under live-like conditions.

\paragraph{Allocation Trajectory Under Unified Execution Interface}

Figure~\ref{fig:Allocation_Trajectory} illustrates the time-varying portfolio weight allocations generated by the strategy during the paper trading window. Rather than emphasizing sector-level performance attribution, the figure highlights how the allocation module produces dynamic weight vectors that are transmitted unchanged through the unified weight-based execution interface.

The observed allocation shifts reflect regime-aware adjustments driven by relative momentum and risk signals. Importantly, no architectural modification was required when transitioning from offline backtesting to broker-level execution, demonstrating structural consistency between research and deployment environments.

\paragraph{Stress Event as Risk-Module Validation}

% The paper-trading window also includes an adverse episode: the portfolio experienced a peak-to-trough drawdown of approximately \textbf{12.2\%} following an extreme move in a leveraged instrument. We treat this episode as a deployment-relevant stress case rather than a performance headline. It highlights the nonlinear exposure characteristics of leveraged products and motivates more adaptive safeguards, such as volatility-regime-aware scaling and instrument-specific exposure caps.

% Importantly, because execution is driven by a unified weight interface, the same post-trade accounting and attribution pipeline can be applied without modifying upstream strategy logic, reinforcing the modularity and diagnosability of the FinRL-X design.

The paper-trading window also includes an adverse episode: the portfolio experienced a peak-to-trough drawdown of approximately \textbf{12.2\%} following an extreme move in a leveraged instrument. We treat this as a deployment-relevant stress case rather than a performance headline, highlighting the nonlinear risk of leveraged products and motivating safeguards such as volatility-aware scaling and instrument-specific exposure caps.

Because execution is driven by a unified weight interface, the same post-trade accounting and attribution pipeline applies without modifying strategy logic, reinforcing the modular and diagnosable design of FinRL-X.

\FloatBarrier
\section{Conclusions}

FinRL-X is a deployment-consistent, modular trading system that unifies data processing, strategy composition, evaluation, and broker execution within a single architecture. By adopting a weight-centric interface, the framework enforces consistent decision semantics across research, backtesting, and live trading, reducing discrepancies between offline evaluation and real-world deployment.

The modular design supports flexible integration of heterogeneous strategies while preserving reproducibility and composability. Empirical evaluation, including broker-integrated paper trading, demonstrates stable execution behavior under realistic conditions. Future work will extend FinRL-X toward broader asset classes and more advanced execution-aware strategies for scalable real-world deployment.

\section*{Acknowledgements}
This work is developed and maintained under the AI4Finance Foundation open-source ecosystem. The AI4Finance Foundation\footnote{\url{https://ai4finance.org}} was founded in 2017 at Columbia University. Some authors contributed to this work while also enrolled as students at Columbia University. FinRL and the FinRL logo are trademarks of FinRL LLC and are used with permission.

%\newpage

\medskip
\small
\bibliographystyle{plain}
\bibliography{ref}

\end{document}